\documentclass{PoS}
\def\be{\begin{equation}}
\def\ee{\end{equation}}
\def\Rir{R_\text{\tiny IR}}
\def\uvec#1{\underline{#1}}

\usepackage{amsmath}

\title{Distribution of the number of particles\\
  in the final state of hadron-nucleus collisions%
  \thanks{This work is supported by the Agence Nationale
    de la Recherche under the project ANR-16-CE31-0019.}}

\ShortTitle{Distribution of the number of particles in $pA$ collisions}

\author{Laura Domin\'e, C\'edric Lorc\'e, \speaker{St\'ephane Munier},
  Simon Pekar\\
        Centre de physique th\'eorique, \'Ecole polytechnique, CNRS, Universit\'e Paris-Saclay, Palaiseau, France}

\abstract{Recently, Liou, Mueller and Munier have argued that
  proton-nucleus collisions at the LHC may give access
  to the full statistics of the event-by-event
  fluctuations of the gluon density in the proton.
  Indeed, the number of particles produced in an event
  in rapidity slices in 
  the fragmentation region of the proton may,
  under some well-defined assumptions, be directly
  related to the number of gluons which have a transverse momentum
  larger than the nuclear saturation scale
  present in the proton at the
  time of the interaction with the nucleus.
  A first calculation of the probability distribution
  of the number of gluons in a hadron was performed,
  using the color dipole model.
  In this talk, we review this proposal, and present
  preliminary numerical calculations which support
  the analytical results obtained so far.
}

\FullConference{XXV International Workshop on Deep-Inelastic Scattering and Related Subjects\\
		3-7 April 2017\\
		University of Birmingham, UK}

\begin{document}

\section{Introduction}

The LHC has opened a new window on hadronic physics by
enabling the measurement of a wealth of new observables in the collisions
of protons and nuclei, at unprecedented center-of-mass energies.
In particular, there is a lot of interest in trying to pin down
higher moments or even the
full distribution of the event-by-event fluctuations of
observables, with a view to characterizing how hadrons
look in individual events. For example,
it has been proposed to use diffractive dissociation processes
to constrain the fluctuations of the shape of the
proton~\cite{Mantysaari:2016jaz}.
This contribution reports on an effort to understand the fluctuations
of the integrated gluon density in the proton.

In the recent paper of Ref.~\cite{Liou:2016mfr}, it has been
proposed to study the
event-by-event fluctuations of the multiplicity of particles produced in
specific regions in rapidity in the final state
of proton-nucleus collisions, arguing that such an observable
would be directly related to the
probability distribution of the gluon density in the proton.
Considering an onium instead of a proton,
a calculation of the probability of particle numbers was attempted,
based on the color dipole model \cite{Mueller:1993rr,Kovchegov:2012mbw}
to represent the QCD evolution, modified in the infrared in order
to mimick confinement.

We shall review this simple picture for particle production
and the corresponding calculation of the multiplicity distribution,
and present preliminary numerical evaluations of the
distribution of the number of particles produced in onium-nucleus
collisions.


\subsection{Picture of particle production in proton-nucleus collisions}

Let us consider the collision of a proton with a large nucleus.
The proton is an initially dilute object, while the
nucleus is a dense object
characterized by the saturation
momentum $Q_A$ in its rest frame.
The physical meaning of $Q_A$ is the following: To
a globally colorless probe whose colored constituents
carry the relative transverse momentum $k\gg Q_A$
(or are separated by the distance $r\sim 1/k\ll 1/Q_A$), the
nucleus appears dilute,
while a probe characterized by $k\ll Q_A$
is absorbed with probability one
in an interaction with this nucleus \cite{Kovchegov:2012mbw}.

The wave function of a nucleus is almost the same as the ground-state
wave function
of nuclear matter, whose density is essentially uniform~\cite{Liou:2016mfr}.
If we look at it from the point of view of a probe at rest
in a frame in which it has the
rapidity $y_0$ (which we shall call the ``$y_0$-frame''\footnote{We
  borrow this terminology and the notations
  from Ref.~\cite{Mueller:2016xti}.}),
then the QCD evolution populates its Fock
state with many gluons, but in an almost deterministic way
since it starts already with a large number of valence quarks.
Hence the state of such a nucleus is still essentially characterized by a
single momentum scale (unfluctuating, i.e. identical in each event),
the saturation momentum $Q_s(y_0)$, larger than $Q_A$
by a factor which grows exponentially with $y_0$.

In the $y_0$-frame,
the proton has the rapidity $Y-y_0$, where $Y$
is the total rapidity of the scattering determined
by the center-of-mass energy. In each event, the proton is found
in a given Fock state
essentially made of many gluons,
produced through QCD evolution.
Since the proton is a dilute object in its rest frame,
when viewed at a nonzero rapidity,
its content is expected to strongly fluctuate from
event to event, unlike the nucleus.
We choose $y_0$ such that there is a significant QCD evolution
in the proton,
but not too large, in such a way that nonlinear saturation effects
can always be neglected.

The effect of the interaction of this proton with the nucleus
is to filter
the gluons which have a transverse momentum less than $Q_s(y_0)$
(thanks to multiple scatterings with
the nucleus which transfer them energy to put them on-shell).
These gluons go to the final state and hadronize.
The other gluons essentially
do not interact and
go back to the vacuum at finite time.
This intuitive picture is backed by QCD
calculations~\cite{Kovchegov:2012mbw}.

Hence the particle multiplicity measured
in the proton fragmentation region in an
event is essentially proportional to
the gluon number density $xG$ integrated
at the scale $Q_s(y_0)$ in the
corresponding realization of the QCD evolution.
More quantitatively, the number $dN/dy_2$ of particles per unit rapidity
measured around the rapidity $y_2=0$ in the
$y_0$-frame reads~\cite{Mueller:2016xti} 
\be
\left.\frac{dN}{dy_2}\right|_{y_2\simeq 0}=xG(x,Q_s^2(y_0)), 
\ee
where $x=e^{-(Y-y_0)}{Q_s(y_0)}/M$
with $M$ the mass of the onium.
Strictly speaking, this formula is established as an equality between
mean quantities (averaged over events),
but we shall assume that it also holds
for each event considered separately.
According to this picture, {\it measuring the distribution
 of the particle number in the final state of $pA$ collisions amounts to
measuring the fluctuations of the gluon number density in the proton.}

So far, experimental and theoretical studies of the
partonic content of the proton have focused on the expectation value
of parton number densities (related
to total cross sections,
for example of deep-inelastic scattering processes)
and their second moments (related to diffractive dissociation).
Hence such an observable in proton-nucleus scattering
would give access to genuinely new information
on the parton densities, namely their full probability
distribution.

We shall now present a QCD calculation of the fluctuations of
the gluon number density in a dilute hadron.


\subsection{How a hadronic state dresses at high energies:
  the color dipole model}

 Instead of addressing proton-nucleus
 collisions, we study the scattering
 of onia (which have the advantage of being color dipoles)
 of initial fixed transverse
 size $\uvec{r}_0$ off large (i.e. infinite transverse extent)
 nuclei.
 An onium is the simplest model of a {\it dilute} hadron, and
 the main features of our calculation
 may go over to the proton case.

In a frame in which the rapidity $y=Y-y_0$ of the onium is large,
the Fock state is a dense system of gluons.
The leading contribution
to the probability of such a state (i.e. the large-$y$ asymptotics)
is obtained by calculating the splitting of the initial quark-antiquark
into a state made of the $q\bar q$ pair supplemented with a soft gluon. The
latter may split further into a denser state populated with softer
and softer gluons. This is essentially the physics captured by the
Balitsky-Fadin-Kuraev-Lipatov (BFKL) equation \cite{Kovchegov:2012mbw}.

\begin{figure}
   \begin{center}
     \includegraphics[width=.9\textwidth]{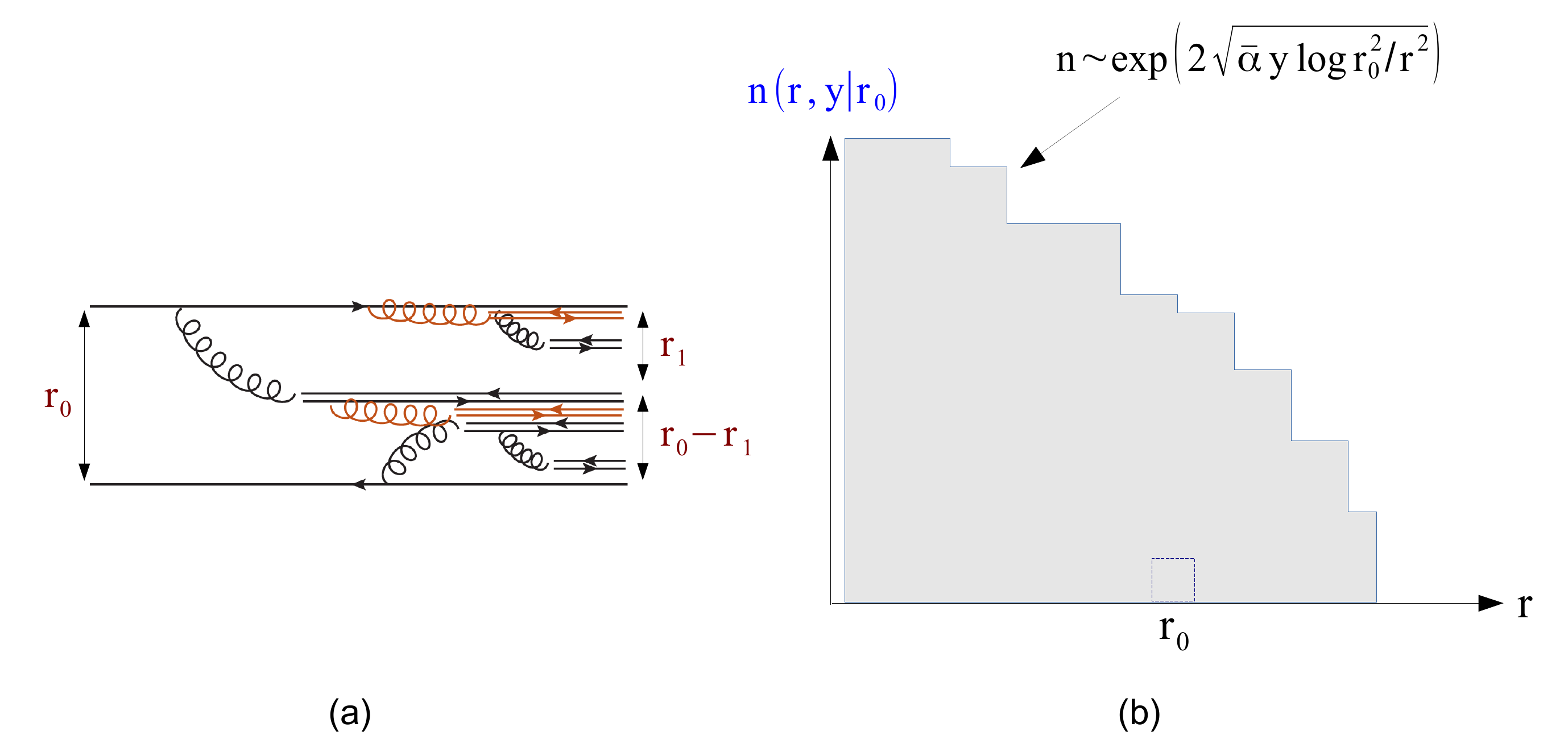}
   \end{center}
   \caption{{\it (a)} Picture of the dressing of an onium of size $r_0$
     boosted to rapidity $y=Y-y_0$ by gluonic fluctuations,
     represented by dipoles in the large-$N_c$ limit.
     Due to the collinear singularity in the splitting probability,
     many small dipoles are produced.
     {\it (b)} Sketch of the number of
     dipoles $n(r,y|r_0)$ of size larger than $r$ in an onium of size $r_0$
     after evolution over the rapidity~$y$
     in a typical
     event, when $y\gg 1/\bar\alpha$, as a function of~$r$
     (logarithmic scale on both axis).
     In the region $r\ll r_0$, $n$ follows the well-known double-logarithmic
     form $n\sim \exp\left(2\sqrt{\bar\alpha y \ln r_0^2/r^2}\right)$,
     which
     is easy to prove for the mean dipole number $\bar n$, but which is
     also true for individual events provided the number of dipoles is
     large enough so that the law of large numbers applies.
\label{fig:dipoles}}
 \end{figure}

Taking the large number-of-color ($N_c$) limit
simplifies the color structure of these
complicated gluonic states, and
leads to the famous color dipole model \cite{Mueller:1993rr}.
The relevant Fock states may
be constructed iteratively from one single elementary process: The splitting
of a dipole of size $\uvec{r}_0$ into two dipoles
of size vectors $\uvec{r}$ and $\uvec{r}-\uvec{r}_0$.
Its rate (as the rapidity increases) is computed in perturbative QCD
and reads
\be
\frac{dP}{dy}=\bar\alpha \frac{d^2\uvec{r}}{2\pi}
\frac{\uvec{r}_0^2}{\uvec{r}^2(\uvec{r}_0-\uvec{r})^2},
\label{eq:dPdy}
\ee
where $\bar\alpha=\alpha_s N_c/\pi$.
This branching process is represented schematically in Fig.~\ref{fig:dipoles},
together with the shape of the dipole distribution obtained in
a typical event.

 Iterating the dipole branching process, we arrive at a random set
 of dipoles. Let us define $n(r,y|r_0)$ the number of dipoles of
 size larger than $r$ in an onium of size $r_0$ at rapidity $y$.
 The density of gluons in the onium carrying the fraction
 $x$ of its momentum and having transverse momentum $k$ is related
 to the number of dipoles $n$ in the following way \cite{Liou:2016mfr}:
 \be
 xG(x,k^2)=\left.\frac{\partial}{\partial y}\right|_{y=\ln(1/x)}
   n(1/k,y|r_0).
 \ee
 Strictly speaking, this relation is established in the double-logarithmic
 approximation, and after averages over the events have been taken.
 Again, we shall assume that it is actually valid for each event.
 
 Non-perturbative physics such as confinement cannot of course
 be captured by
 the original color dipole model. Roughly speaking, confinement restricts
 the size of color dipoles to be less than the typical hadronic size
 $1/\Lambda_\text{QCD}$. We shall model it 
 by multiplying the dipole splitting rate 
 by the ad hoc cutoff function $\Theta$:
 \be
 \left.\frac{dP}{dy}\right|_\text{Eq.~(\ref{eq:dPdy})}
 \longrightarrow\ \frac{dP}{dy}
 =\bar\alpha \frac{d^2\uvec{r}}{2\pi}
 \frac{\uvec{r}_0^2}{\uvec{r}^2(\uvec{r}_0-\uvec{r})^2}
 \Theta(\uvec{r},\uvec{r}_0-\uvec{r};\Rir),
 \ee
chosen in such a way that $\Theta(\uvec{r},\uvec{r}_0-\uvec{r};\Rir)$
 tends to zero when $r\gg\Rir$
 or $|\uvec{r}_0-\uvec{r}|\gg\Rir$ and to one when both $r$ and
 $|\uvec{r}_0-\uvec{r}|$ are
 much less than $\Rir$.
 

\section{Probability distribution of the particle multiplicity}

\subsection{Analytical insights}

Let us define $P_n(r,y)$ the probability
to have $n$ dipoles of size larger than $1/Q_s(y_0)$ in
the Fock state at rapidity $y$.
It is not difficult to write down
a system of evolution equations for $P_n$:
\be
\frac{\partial P_n(r_0,y)}{\partial y}=
\bar\alpha\int
\frac{d^2\uvec{r}}{2\pi}\frac{\uvec{r}_0^2}{\uvec{r}^2(\uvec{r}_0-\uvec{r})^2}
\Theta(\uvec{r},\uvec{r}_0-\uvec{r};\Rir)
\left[
  \sum_{m=1}^{n-1}P_m(r,y)P_{n-m}(|\uvec{r}_0-\uvec{r}|,y)-P_n(r_0,y)
  \right].
\label{eq:eqPn}
\ee
This infinite set of equations may be summarized by writing
the evolution equation for the generating functional of
the factorial moments of $P_n$, which turns out to coincide with the
Balitsky-Kovchegov (BK) equation~\cite{Kovchegov:2012mbw}.
Hence, solving~(\ref{eq:eqPn}) is tantamount to solving the BK
equation, which so far has not been achieved.

The analytical and
numerical study of the fluctuations of the number of dipoles was pioneered
in Ref.~\cite{Salam:1995zd}
in the context of the color dipole model without
an infrared cutoff (i.e. for $\Theta=1$).
In this case, the probability $P_n$
was found to exhibit a fat tail
\be
P_n\underset{n\gg\bar n}{\sim}
\exp\left(
-\frac{\ln^2n}{4\bar\alpha y}
\right).
\ee
The large-multiplicity events are characterized by
the presence of a large dipole,
produced at the very beginning of the QCD evolution.
Its size may become arbitrarily large as one
selects events which have a larger and larger occupancy.

When confinement is accounted for, such large dipoles may not be
produced. This results in a drastic damping of the tail of the
probability distribution, which becomes exponential:
\be
P_n\underset{n\gg\bar n}{\sim}e^{-n/(c\bar n)},
\label{eq:Pn}
\ee
where $\bar{n}$ is the mean dipole number (i.e. averaged over
events)
and $c$ is a constant a priori of order~1, but which we have not
been able to determine analytically.

The detailed calculation which leads
to Eq.~(\ref{eq:Pn}) can be found in Ref.~\cite{Liou:2016mfr,toappear}.
Since it relies on an Ansatz, numerical checks are required.


\subsection{Numerical checks}

\begin{figure}
   \begin{center}
     \includegraphics[width=\textwidth]{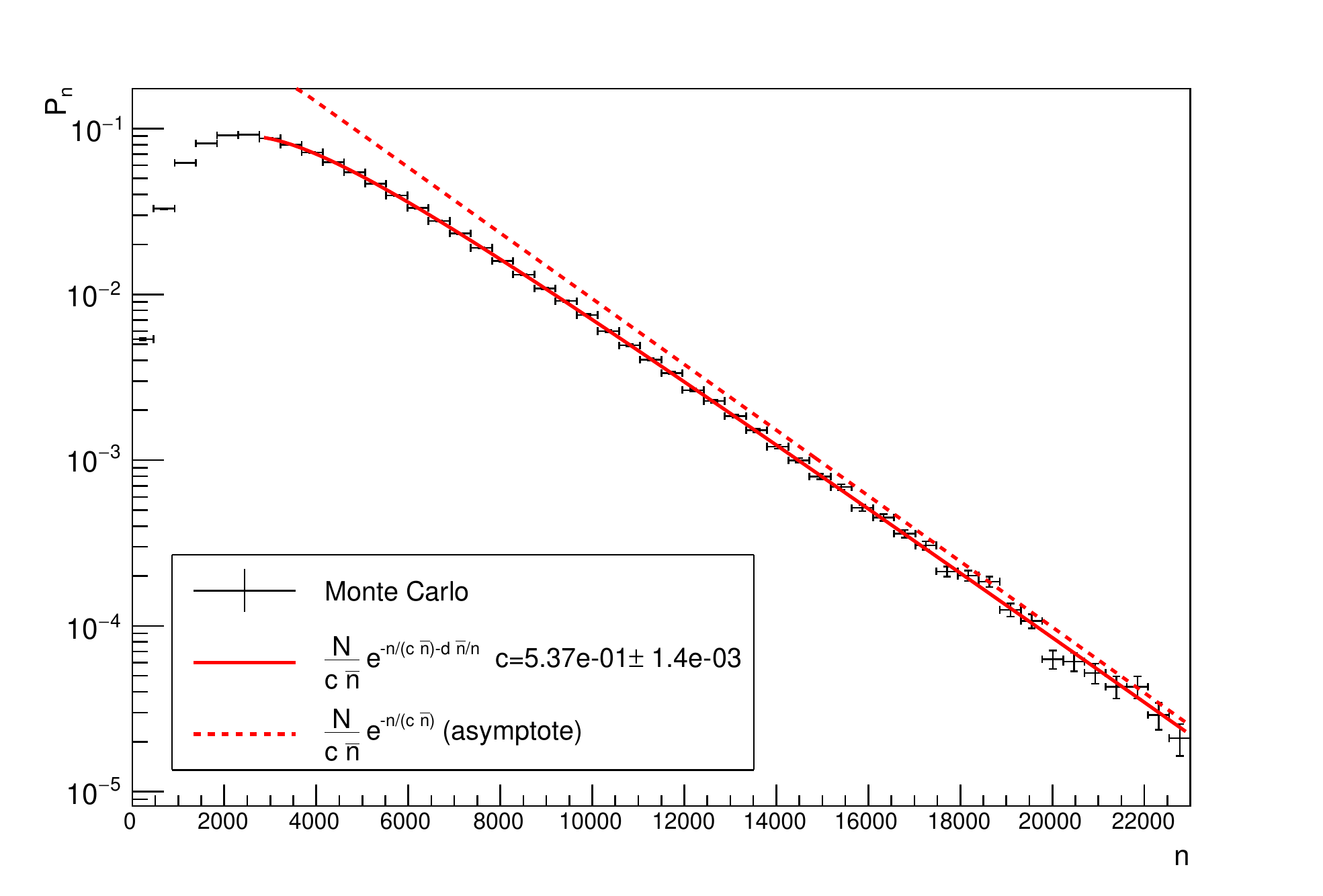}
   \end{center}
   \caption{{\it Points:} Numerical data for $P_n$ for a particular
     cutoff function $\Theta$ and set of parameters (see the main text for details).
     {\it Full line:} Graph of the 
     function $N/(c\bar n)e^{-n/(c\bar{n})-d\bar{n}/n}$, where
     the parameters $N$, $c$ and $d$ are determined from the least square
     method ($\chi^2/d.o.f.\simeq 38/41$ when
     $N\simeq 0.9$, $c\simeq 0.537$ and $d\simeq 0.71$).
     Note that we have added the arbitrary term $-d\times\bar{n}/n$ in the
     exponent to effectively take into account subasymptotic contributions
     which, so far, we have not been able to determine analytically.
     {\it Dashed line:} Large-$n$ asymptote.
\label{fig:Pn}}
 \end{figure}

We have implemented numerically the color dipole model
in the form of a Monte Carlo event generator
along the lines
of Ref.~\cite{Salam:1996nb}, but with a modified BFKL kernel 
which simulates the effect of confinement
through the cutoff function~$\Theta$.
We present in Fig.~\ref{fig:Pn} the numerical results
for $P_n$ for the following choice of this function:
\be
\Theta(\uvec{r}_1,\uvec{r}_2;\Rir)=\exp\left(
-\frac{r_1^2+r_2^2}{2\Rir^2}
\right),
\ee
and setting the parameters to the values $\bar\alpha y=3$,
$Q_A=50/r_0$, $\Rir=2r_0$.
(This Gaussian form was part of the Ansatz in Ref.~\cite{Liou:2016mfr}).
 Figure~\ref{fig:Pn} shows that the asymptotics~(\ref{eq:Pn})
 with a constant $c$ in the range $0.5$~--~$0.6$
 are fully consistent with the numerical calculations.

 We are now aiming at investigating
 how the decay of $P_n$ at large $n$ depends on the infrared
 sector, i.e. how confinement is enforced onto the perturbative
 QCD calculation. Preliminary results seem to indicate that the
 dependence on the form of the cutoff function $\Theta$
 is rather mild~\cite{toappear}. 
 

\end{document}